\documentclass[aps,prd,twocolumn,amsmath,amssymb,showpacs]{revtex4}

\usepackage{graphicx}
\usepackage{dcolumn}
\usepackage{bm}

\begin{document}

\title{Quasinormal modes of the BTZ black hole are generated by surface waves
supported by its boundary at infinity}

\author{Yves D\'ecanini}
\email{decanini@univ-corse.fr}
\author{Antoine Folacci}
\email{folacci@univ-corse.fr} \affiliation{
UMR CNRS 6134 SPE, Equipe Physique Semi-Classique (et) de la Mati\`ere Condens\'ee \\
Universit\'e de Corse, Facult\'e des Sciences, BP 52, 20250 Corte,
France}

\date{\today}

\begin{abstract}
\bigskip
We develop the complex angular momentum method in the context of the
BTZ black hole physics. This is achieved by extending a formalism
introduced a long time ago by Arnold Sommerfeld, which allows us to
define and use the Regge pole concept in a framework where the
notion of an $S$ matrix does not exist. The Regge poles of the BTZ
black hole are exactly obtained and from the associated Regge
trajectories we determine its quasinormal mode complex frequencies.
Furthermore, our approach permits us to physically interpret them:
they appear as Breit-Wigner-type resonances generated by surface
waves supported by the black hole boundary at infinity which acts as
a photon sphere.
\bigskip\bigskip
\end{abstract}


\maketitle

\section{Introduction}

By using complex angular momentum (CAM) techniques, we showed some
years ago that the quasinormal mode (QNM) complex frequencies of the
Schwarzschild black hole of mass $M$ are Breit-Wigner-type
resonances generated by a family of ``surface waves" lying on its
photon sphere at $r=3M$ \cite{DecaniniFJ_cam_bh}. More precisely, by
noting that each surface wave is associated with a Regge pole of the
$S$ matrix of the Schwarzschild black hole
\cite{Andersson1,Andersson2,DecaniniFJ_cam_bh}, we have been able to
construct the spectrum of the QNM complex frequencies from the Regge
trajectories, i.e., from the curves traced out in the CAM plane by
the Regge poles as a function of the frequency. In this way, we have
established, on a ``rigorous" basis, an appealing and physically
intuitive interpretation of the Schwarzschild black hole QNMs
suggested by Goebel in 1972 \cite{Goebel}, i.e., that they could be
interpreted in terms of gravitational waves in spiral orbits close
to the unstable circular photon orbit at $r=3M$ which decay by
radiating away energy (see also
Refs.~\cite{FerrariMashhoon84,Mashhoon85,Stewart89,
AnderssonOnozawa96} for alternative implementations of the Goebel
interpretation).

Now, it is natural to use the CAM approach in order to physically
understand the resonant aspects of more general black holes. We
believe that, {\it mutatis mutandis}, it should be fairly easy to
generalize the CAM analysis developed in
Refs.~\cite{Andersson1,Andersson2,DecaniniFJ_cam_bh} for all the
asymptotically flat black holes with a photon sphere. By contrast,
for black holes immersed in a non-asymptotically flat background,
i.e., in a framework where the notion of an $S$ matrix does not
exist, we can expect to encounter new difficulties. However, because
such black holes play a central role in the context of superstring
theory and quantum gravity, their CAM analysis seems to be an
interesting and important task which could, in particular, shed
light on AdS/CFT correspondence and holography from a new point of
view.

In the present paper, our aim is more modest even if we try to make
some steps in this direction. We shall consider the most simple
black hole immersed in an asymptotically anti-de Sitter space-time,
namely the (2+1)-dimensional Ba{\~n}ados-Teitelboim-Zanelli (BTZ)
black hole \cite{btz}. It seems to us very interesting in order to
test our approach because, in that particular space-time, the wave
equation can be solved exactly
\cite{Ghoroku1994ij,Ichinose1994rg,CardosoLemos2001,Birmingham2001}.
We shall revisit the QNM problem for this black hole from the point
of view of the CAM approach in order to analyze it semiclassically,
i.e., in term of surface waves. In Sec.~II, we shall define the
Regge poles and the associated Regge modes for a massless scalar
field defined on the BTZ black hole by extending a formalism
introduced a long time ago by Sommerfeld \cite{Sommerfeld49} as an
alternative to the usual Watson approach of scattering
\cite{Watson18}. It permits us to use the Regge pole concept in a
framework where the notion of an $S$ matrix does not exist. The
Regge poles and the Regge modes of the BTZ black hole are exactly
obtained and physically interpreted in terms of surface waves
supported by its boundary, this boundary furthermore playing the
role of a photon sphere. In Sec.~III, we shall first construct from
the Regge modes the diffractive part of the Feynman propagator
associated with the scalar field. We shall then show that the poles
of its temporal Fourier transform are the QNM complex frequencies of
the BTZ black hole and prove that they are generated by the surface
waves supported by the black hole boundary. In a short conclusion,
we shall make some remarks concerning the AdS/CFT correspondence and
the possibility to consider black hole photon spheres as holographic
screens.

\section{Regge modes of the BTZ black hole}

\subsection{Quasinormal modes and Regge modes of the BTZ black hole}

The metric of the spinless BTZ black hole with mass $M>0$
``immersed" into $\mathrm{AdS}_3$ with length scale $\ell$ is given
by
\begin{equation} \label{BTZmetric}
ds^2=-\bigg(\frac{r^2-r_h^2}{\ell^2}\bigg)dt^2
+\bigg(\frac{r^2-r_h^2}{\ell ^2}\bigg)^{-1}dr^2+r^2d\phi^2
\end{equation}
where $t\in ]-\infty,+\infty[$, $r>r_h$ and $\phi$ has period
$2\pi$. Here $r_h=\ell \sqrt{M}$ denotes the horizon radius of the
BTZ black hole. Propagating on this gravitational background, we
consider a massless minimally coupled scalar field $\Phi$ solution
of the wave equation
\begin{equation}\label{WEq1}
\Box \Phi =0.
\end{equation}
By inserting the metric (\ref{BTZmetric}) into (\ref{WEq1}), the
wave equation provides
\begin{eqnarray}\label{WEq2}
& & -\bigg(\frac{r^2-r_h^2}{\ell^2}\bigg)^{-1}\frac{\partial ^2
\Phi}{\partial
t^2}+\left(\frac{r^2-r_h^2}{\ell^2}\right)\frac{\partial ^2
\Phi}{\partial r^2} \nonumber\\ & & \qquad\qquad
+\bigg(\frac{3r^2-r_h^2}{r\ell^2}\bigg)\frac{\partial \Phi}{\partial
r} +\frac{1}{r^2}\frac{\partial ^2 \Phi}{\partial \phi^2} =0
\end{eqnarray}
and we can look for its solutions by separation of variables or,
more precisely, by using the ansatz
\begin{equation}\label{ModeSol_1}
\Phi_{\nu,\omega}(t,r,\phi)=\frac{1}{\sqrt{r}}f_{\nu,\omega}(r)e^{i(\nu
\phi -\omega t)}.
\end{equation}
The radial equation satisfied by $f_{\nu,\omega}(r)$ takes the form
\begin{eqnarray}\label{WEq3}
& & \left(\frac{r^2-r_h^2}{\ell^2}\right)\frac{d^2
f_{\nu,\omega}(r)}{dr^2} + \frac{2r}{\ell^2} \frac{d\,
f_{\nu,\omega}(r)}{d r} \nonumber\\ & & \quad
+\left[\bigg(\frac{r^2-r_h^2}{\ell^2}\bigg)^{-1}\omega^2-\frac{\nu^2}{r^2}
 -\frac{r_h^2}{4\ell^2 r^2} -\frac{3}{4\ell^2} \right]f_{\nu,\omega}(r)
 \nonumber\\ & & \quad =0.
\end{eqnarray}
This differential equation can be solved exactly (see
\cite{Ghoroku1994ij,Ichinose1994rg,CardosoLemos2001,Birmingham2001})
in terms of hypergeometric functions \cite{AS65}. We then obtain for
the general form of its solution
\begin{widetext}
\begin{eqnarray}\label{ModeSol_GenForm}
&& f_{\nu,\omega}(r)=  A_{\nu,\omega}\left(\frac{r}{r_h}
\right)^{\frac{1}{2}}\left(1-\frac{r_h^2}{r^2}
\right)^{-i\frac{\ell^2\omega}{2 r_h}}
F\left[-\frac{i\ell}{2r_h}(\ell\omega-\nu),-\frac{i\ell}{2r_h}(\ell\omega+\nu)
;
1-i\frac{\ell^2\omega}{r_h} ;1-\frac{r_h^2}{r^2}\right]   \nonumber\\
& & \qquad   + B_{\nu,\omega}\left(\frac{r}{r_h}
\right)^{\frac{1}{2}}\left(1-\frac{r_h^2}{r^2}
\right)^{+i\frac{\ell^2\omega}{2 r_h}}
F\left[+\frac{i\ell}{2r_h}(\ell\omega-\nu),+\frac{i\ell}{2r_h}(\ell\omega+\nu)
;1+i\frac{\ell^2\omega}{r_h} ;1-\frac{r_h^2}{r^2}\right]
\end{eqnarray}
\end{widetext}
where $A_{\nu,\omega}$ and $B_{\nu,\omega}$ are arbitrary complex
constants.

From now on, we consider more particularly the mode solutions
defined by (\ref{ModeSol_1}) and (\ref{ModeSol_GenForm}) which
correspond to the so-called QNMs and the so-called Regge modes. They
are both defined as mode solutions which are purely ingoing at the
horizon (i.e., for $r=r_h$) and vanish at infinity (i.e., for
$r=+\infty$). The first condition selects $B_{\nu,\omega}=0$. By
noting that $F(a,b;c;1)=[\Gamma(c)\Gamma(c-a-b)
]/[\Gamma(c-a)\Gamma(c-b) ]$ if $c\not=0,-1,-2, \dots$ and
$\mathrm{Re}(c-a-b)>0$, the second one imposes
\begin{subequations}\label{Cond_QNM_RM}
\begin{equation}\label{Cond_QNM_RMa}
1- \frac{i\ell}{2r_h}(\ell\omega-\nu)= -n  \quad \mathrm{with} \quad
n \in \mathbf{N}
\end{equation}
or
\begin{equation}\label{Cond_QNM_RMb}
1- \frac{i\ell}{2r_h}(\ell\omega+\nu)= -n  \quad \mathrm{with} \quad
n \in \mathbf{N}.
\end{equation}
\end{subequations}

Let us first consider the QNMs. Because they are periodic in $\phi$,
we must have $\nu=m\in \mathrm{Z}$. They are therefore only defined
for the discrete complex values of the frequency $\omega$ (see also
Refs.~\cite{CardosoLemos2001,Birmingham2001})
\begin{equation}\label{freqcomp_QNM}
\omega_{mn}^\pm=\pm\frac{m}{\ell}-i\frac{2r_h}{\ell^2}(n+1) \quad
\mathrm{with} \,\, m \in \mathbf{Z} \,\, \mathrm{and} \,\, n\in
\mathbf{N}
\end{equation}
[here, the plus sign correspond to (\ref{Cond_QNM_RMa}) and the
minus one to (\ref{Cond_QNM_RMb})] and they are given by
\begin{equation}\label{ModeSol_QNM_1}
\Phi^{\pm}_{m n}(t,r,\phi)=\frac{1}{\sqrt{r}}f^{\pm}_{m n}(r)e^{i(m
\phi -\omega^{\pm}_{m n} t)}
\end{equation}
with
\begin{eqnarray}\label{ModeSol_QNM_2}
&& f^{\pm}_{m n}(r)=  A^{\pm}_{m n}  \left(\frac{r}{r_h}
\right)^{\frac{1}{2}}\left(1-\frac{r_h^2}{r^2}
\right)^{-i\frac{\ell^2 \omega^{\pm}_{m n}}{2 r_h}} \nonumber\\
& &   \times F\left[-n-1,-\frac{i\ell}{2r_h}(\ell\omega^{\pm}_{m n}
\pm m)
; 1-i\frac{\ell^2\omega^{\pm}_{m n}}{r_h} ;1-\frac{r_h^2}{r^2}\right].  \nonumber\\
& &
\end{eqnarray}
It should be noted that $\forall m\in \mathbf{Z}$,
$\omega_{-mn}^+=\omega_{mn}^-$ but $\Phi^{+}_{-m n}(t,r,\phi) \not=
\Phi^{-}_{m n}(t,r,\phi)$. As a consequence, each QNM complex
frequency is two-fold degenerated.

Let us now consider the Regge modes. They are defined on the
covering space of the BTZ black hole obtained by relaxing the
condition of periodicity in the coordinate $\phi$, i.e., by
considering that $\phi \in ]-\infty,+\infty[$, and by furthermore
assuming that $\omega >0$. The covering space of the BTZ black hole
considered here is, in fact, nothing other than one of the 12 charts
which permits us to provide a global covering of $\mathrm{AdS}_3$ in
BTZ coordinates \cite{btzH}. However, it should be noted that we do
not work on $\mathrm{AdS}_3$: relaxing the condition of periodicity
in the coordinate $\phi$ is just a trick which will permit us to
consider multivalued mode solutions of the wave equation
(\ref{WEq1}) which describe surface waves propagating around the BTZ
black hole and to take into account their multiple circumnavigations
as well as the associated radiation damping due to their
attenuations. Such a trick has been invented by Sommerfeld in order
to analyze scattering by spheres and to emphasize the role of
surface waves (see Ref.~\cite{Sommerfeld49} and, more particularly,
Appendix II of Chapter V as well as the appendix of Chapter VI). The
Sommerfeld approach is an alternative to the usual CAM approach of
scattering developed by Watson \cite{Watson18}. It allows us to use
the Regge pole concept in a framework where the notion of an $S$
matrix does not exist, and therefore to extend our CAM analysis of
the Schwarzschild black hole \cite{DecaniniFJ_cam_bh} to the BTZ
one.

The Regge modes of the BTZ black hole are only defined for
particular complex values of the parameter $\nu$ which we shall call
Regge poles even if they are not the poles of an $S$ matrix. These
Regge poles are exactly given by
\begin{equation} \label{ReggeP}
\nu_{n}^\pm(\omega) =\pm \omega \ell \pm i\frac{2r_h}{\ell}(n+1)
\quad \mathrm{with} \,\, \omega>0  \,\, \mathrm{and} \,\, n\in
\mathbf{N}
\end{equation}
[here, the plus sign correspond to (\ref{Cond_QNM_RMa}) and the
minus one to (\ref{Cond_QNM_RMb})] and the corresponding Regge modes
are given by
\begin{subequations}\label{ReggeM_1}
\begin{eqnarray}
& & \Phi_{\nu_{n}(\omega)}^\pm
(t,r,\phi)=\frac{1}{\sqrt{r}}f_{\nu_{n}(\omega)}^\pm (r)
e^{i[\nu_{n}^\pm(\omega) \phi -\omega t]}  \label{ReggeM_1a}\\
& & \phantom{\Phi_{\nu_{n}(\omega)}^\pm
(t,r,\phi)}=\frac{1}{\sqrt{r}}f_{\nu_{n}(\omega)}^\pm (r) e^{\mp
\frac{2r_h}{\ell}(n+1)\phi} e^{i[\pm \omega \ell \phi -\omega t]}
\nonumber \\
&& \label{ReggeM_1b}
\end{eqnarray}
\end{subequations}
with
\begin{eqnarray}\label{ReggeM_2}
&& f_{\nu_{n}(\omega)}^\pm(r)=  A_{\nu_{n}(\omega)}^\pm
\left(\frac{r}{r_h} \right)^{\frac{1}{2}}\left(1-\frac{r_h^2}{r^2}
\right)^{-i\frac{\ell^2 \omega}{2 r_h}} \nonumber\\
& &   \times F\left[-n-1,n+1-i\frac{\ell^2\omega}{r_h}
; 1-i\frac{\ell^2\omega}{r_h} ;1-\frac{r_h^2}{r^2}\right].  \nonumber\\
& &
\end{eqnarray}
It should be noted that, even if the radial parts of
$\Phi_{\nu_{n}(\omega)}^+ $ and $\Phi_{\nu_{n}(\omega)}^- $ are
identical, these mode solutions are different. In fact, because we
have $\nu_{n}^+(\omega)=-\nu_{n}^-(\omega)$, they respectively
describe waves with identical properties but propagating
counterclockwise and clockwise with an exponential decay around the
BTZ black hole, with $|\mathrm{Re} \, \nu_{n}^\pm(\omega)|=\omega
\ell$ representing their azimuthal propagation constant and
$|\mathrm{Im} \, \nu_{n}^\pm(\omega)|=(2r_h/\ell)(n+1)$ their
damping constant.

\subsection{More on the physical interpretation of the Regge modes}

The Regge poles and the Regge modes obtained in the previous
subsection can be semiclassically interpreted in terms of surface
waves supported by the boundary at infinity of the BTZ black hole,
this boundary furthermore playing the role of a photon sphere. We
shall now discuss more precisely these two important results which
permit us to establish some interesting analogies between the BTZ
and the Schwarzschild black holes.

The propagative behavior in $\exp (i[\mathrm{Re} \,
\nu_{n}^\pm(\omega) \phi -\omega t])=\exp (i[\pm \omega \ell \phi
-\omega t])$ of the Regge modes (\ref{ReggeM_1})-(\ref{ReggeM_2})
permits us to note that the waves they describe circle the BTZ black
hole in time
\begin{equation} \label{ReggeP_interp1}
T = \frac{2\pi}{\omega} |\mathrm{Re} \, \nu_{n}^\pm(\omega)|=2\pi
\ell.
\end{equation}
Furthermore, a scalar photon (associated with the massless scalar
field $\Phi$) on the circular orbit with constant radius $R$ takes
the time
\begin{equation} \label{ReggeP_interp2}
T'= \frac{2\pi R}{\sqrt{\frac{R^2-r_h^2}{\ell^2}}}= \frac{2\pi
R}{\sqrt{\frac{R^2}{\ell^2}-M}}
\end{equation}
to circle the BTZ black hole. Such result can be easily found by
solving $ds^2 = 0$ with $ds^2$ given by (\ref{BTZmetric}), i.e. by
integrating the equation of a circular null geodesic. By equating
$T$ and $T'$, we obtain that necessarily $R \to +\infty$. In other
words, the circular orbit of the scalar photon lies on the BTZ black
hole boundary and we can consider that the set of Regge modes
constitutes a family indexed by $n\in \mathbf{N}$ of surface waves
supported by this boundary.

Let us now consider the photon sphere of the BTZ black hole. We
recall that, for a static spherically symmetric black hole of
dimension $d$ with metric of the form
\begin{equation}\label{ds2_gen}
ds^2=-g_{tt}(r)dt^2+g_{rr}(r)dr^2+r^2 d\Omega^2_{d-2},
\end{equation}
the photon sphere is defined by the greater positive solution of the
equation
\begin{equation}\label{Eq_PhotSphere}
\frac{g'_{tt}(r)}{g_{tt}(r)}=\frac{2}{r}
\end{equation}
[see Ref.~\cite{ClaudelVirbhadraEllis2001} for a general definition
and an extension of the photon sphere concept in an arbitrary
space-time and Eq.~(54) of this paper]. For the BTZ black hole,
(\ref{Eq_PhotSphere}) admits a unique formal solution for $r\to
+\infty$. In that sense, the boundary of the BTZ black hole can be
formally considered as its photon sphere which then acts as the
support of the Regge surface waves
(\ref{ReggeM_1})-(\ref{ReggeM_2}).

\section{Diffracted Feynman propagator and quasinormal frequencies}

The Feynman propagator associated with the scalar field $\Phi$ as
well as the corresponding retarded and advanced Green functions
satisfy the wave equation
\begin{equation}\label{WEQ_GF_GR_GA_1}
\Box_x G(x,x')=- \delta^3(x,x')
\end{equation}
which, in the BTZ black hole space-time defined by
(\ref{BTZmetric}), takes the form
\begin{eqnarray}\label{WEQ_GF_GR_GA_2}
& & \left[ -\bigg(\frac{r^2-r_h^2}{\ell^2}\bigg)^{-1}\frac{\partial
^2 }{\partial t^2}+
\left(\frac{r^2-r_h^2}{\ell^2}\right)\frac{\partial ^2 }{\partial
r^2} \right. \nonumber\\ & & \qquad \left.
+\bigg(\frac{3r^2-r_h^2}{r\ell^2}\bigg)\frac{\partial }{\partial r}
+\frac{1}{r^2}\frac{\partial ^2 }{\partial \phi^2}
\right]G(t,r,\phi;t',r',\phi') \nonumber\\ & & \qquad\qquad\qquad
=-\frac{1}{r}\delta(t-t') \delta(r-r')\delta(\phi-\phi').
\end{eqnarray}
All these Green functions can be constructed by Fourier transform
from the Green function $G_\omega(r,\phi;r',\phi')$ with $\omega>0$
defined as the symmetric solution of the Helmholtz-type equation
\begin{eqnarray}\label{Helmhotz}
& & \left[ \left(\frac{r^2-r_h^2}{\ell^2}\right)\frac{\partial ^2
}{\partial r^2}
+\bigg(\frac{3r^2-r_h^2}{r\ell^2}\bigg)\frac{\partial }{\partial r}
\right. \nonumber\\ & & \qquad\qquad \left.
+\frac{1}{r^2}\frac{\partial ^2 }{\partial \phi^2}
+\bigg(\frac{r^2-r_h^2}{\ell^2}\bigg)^{-1}\omega^2
\right]G_\omega(r,\phi;r',\phi') \nonumber\\ & &
\qquad\qquad\qquad\qquad =-\frac{1}{2\pi r}
\delta(r-r')\delta(\phi-\phi')
\end{eqnarray}
which vanishes at infinity. For example, we have for the Feynman
propagator
\begin{eqnarray}\label{GF et Gom}
& & G_F(t,r,\phi;t',r',\phi')=\frac{1}{2\pi} \int_{-\infty}^{+\infty}
[\Theta (+\omega) G_{+\omega}(r,\phi;r',\phi')  \nonumber \\
& & \qquad\qquad + \Theta (-\omega) G_{-\omega}
(r,\phi;r',\phi')]e^{i[\omega(t-t')]} \, d\omega.
\end{eqnarray}
Here, $\Theta $ denotes the Heaviside step function and it should be
noted that, in order to construct the Feynman propagator
$G_F(t,r,\phi;t',r',\phi')$, we need the Green function
$G_\omega(r,\phi;r',\phi')$ also for $\omega<0$. In fact, as we
shall see later, we will also need $G_\omega(r,\phi;r',\phi')$ in
the full complex $\omega$-plane in order to discuss the resonant
aspects of the BTZ black hole. It can be obtained by analytic
continuation from its expression for $\omega >0$.

From now on, we shall mainly focus our attention on the Green
function $G_\omega(r,\phi;r',\phi')$. Following Sommerfeld
\cite{Sommerfeld49} (see also Sec.~2.1 and Appendix A.1 of
Ref.~\cite{AFG2000a} for a pedagogical introduction to the
Sommerfeld method and the full paper for its application to
cylinders), we construct $G_\omega(r,\phi;r',\phi')$ from the Regge
modes defined in the previous section or, more precisely, we seek it
in the form
\begin{eqnarray}\label{G_helm_diff}
&&
G_\omega^{d}(r,\phi;r',\phi')=\frac{1}{\sqrt{r}}\sum_{n=0}^{+\infty}\,
[f_{\nu_{n}^{+}(\omega)}(r)  V^+_{n,\omega}(\phi  ;  r',\phi')
\nonumber
\\
& & \qquad \qquad +f_{\nu_{n}^{-}(\omega)}(r)  V^-_{n,\omega}(\phi ;
r',\phi')].
\end{eqnarray}
Here we consider that the functions $f_{\nu_{n}^\pm(\omega)}(r)$ are
given by (\ref{ReggeM_2}) with $A_{\nu_{n}(\omega)}^\pm=1$ and we
assume that the functions $V^\pm_{n,\omega}(\phi ; r',\phi')$ are
such that $G_\omega^{d}(r,\phi;r',\phi')$ is a solution of
(\ref{Helmhotz}) symmetric under the exchange $(r,\phi)
\leftrightarrow (r',\phi')$. Furthermore, by working in Sec.~II with
the Regge modes, we have deferred the imposition of periodicity in
the coordinate $\phi$ on the solutions of the wave equation but now
we shall impose this condition on (\ref{G_helm_diff}) and therefore
on (\ref{GF et Gom}).

It is important to note that the Green function constructed by
inserting (\ref{G_helm_diff}) into (\ref{GF et Gom}) differs from
the Feynman propagator which could be obtained from the exact
normalized mode solutions of the wave equation (\ref{WEq1}). Indeed,
the Regge modes do not constitute a complete system of solution of
this equation. In fact, the Sommerfeld method permits us to only
consider that part of the exact Feynman propagator which describe
more particularly ``diffraction" by the BTZ black hole. Such a
result was noted by Sommerfeld for its analysis of scattering by
spheres and remains valid in the present context. This drawback is
not too serious because, in fact, it is the diffractive part of the
Feynman propagator which contains all the information about the
resonant aspects of the problem.

By inserting (\ref{G_helm_diff}) into (\ref{Helmhotz}) and by using
(\ref{WEq3}) with $\nu=\nu_{n}^\pm(\omega)$, we find that the
functions $V^\pm_{n,\omega}(\phi ; r',\phi')$ must satisfy
\begin{eqnarray}\label{eq des Vn_1}
&& \frac{1}{r^2} \sum_{n=0}^{+\infty}\,
\left[f_{\nu_{n}^{+}(\omega)}(r) \left(\frac{\partial^2}{\partial
\phi^2}+{\nu_{n}^{+}(\omega)}^2\right)V^+_{n,\omega}(\phi ;
r',\phi') \right. \nonumber
\\
& & \qquad  \left. +f_{\nu_{n}^{-}(\omega)}(r)
\left(\frac{\partial^2}{\partial
\phi^2}+{\nu_{n}^{-}(\omega)}^2\right)V^-_{n,\omega}(\phi ;
r',\phi') \right] \nonumber \\
& & \qquad \qquad = -\frac{1}{2\pi \sqrt{r}}
\delta(r-r')\delta(\phi-\phi').
\end{eqnarray}
We then multiply (\ref{eq des Vn_1}) by $f_{\nu_{p}^\pm(\omega)}(r)$
and integrate over the radial domain $r\in [r_h,+\infty[$ which
contains $r'$. The orthonormalization relation
\begin{equation}\label{ortho_RM_1}
\int_{r_h}^{+\infty} \frac{1}{r^2}
f_{\nu_{n}^\pm(\omega)}(r)f_{\nu_{p}^\pm(\omega)}(r) \, dr= \pm
N^\pm_n(\omega) \delta_{np}
\end{equation}
with
\begin{eqnarray}\label{ortho_RM_2}
& & N^\pm_n(\omega) =i \, \frac{n! (n+1)!}{2\ell \,
{\nu_{n}^\pm(\omega)}}\nonumber
\\
&& \qquad \times
\frac{\left[\Gamma\left(1-i\frac{\ell^2\omega}{r_h}\right)\right]^2}
{\Gamma\left(n+1-i\frac{\ell^2\omega}{r_h}\right)
\Gamma\left(n+2-i\frac{\ell^2\omega}{r_h}\right)}
\end{eqnarray}
for the Regge radial modes which can be obtained from (\ref{WEq3})
by generalizing, {\it mutatis mutandis}, the calculation displayed
in Appendix A.1 of Ref.~\cite{AFG2000a}, permits us to write
\begin{eqnarray}\label{Eq_des_Vn_2}
& & \left(\frac{\partial^2}{\partial
\phi^2}+{\nu_{n}^\pm(\omega)}^2\right)V^\pm_{n,\omega}(\phi ;
r',\phi')= \nonumber \\
& & \qquad \mp \frac{1}{2\pi
N^\pm_n(\omega)}\frac{f_{\nu_{n}^\pm(\omega)}(r')}{\sqrt{r'}}
\delta(\phi-\phi').
\end{eqnarray}
The general solution of (\ref{Eq_des_Vn_2}) can be sought in the
form
\begin{eqnarray}\label{Exp_des_Vn_1}
& & V^\pm_{n,\omega}(\phi ; r',\phi')= \Theta(\phi' -\phi) [ {\cal
A}^\pm_{n,\omega}(r',\phi')
\cos ( {\nu_{n}^\pm(\omega)}  \phi)  \nonumber \\
& & \qquad + {\cal B}^\pm_{n,\omega}(r',\phi') \sin (
{\nu_{n}^\pm(\omega)}\phi) ] \nonumber \\
& & \phantom{V^\pm_{n,\omega}(\phi ; r',\phi')} + \Theta(\phi
-\phi') [ {\cal C}^\pm_{n,\omega}(r',\phi')
\cos ( {\nu_{n}^\pm(\omega)}  \phi)  \nonumber \\
& & \qquad + {\cal D}^\pm_{n,\omega}(r',\phi') \sin (
{\nu_{n}^\pm(\omega)}\phi) ].
\end{eqnarray}
Then, by inserting (\ref{Exp_des_Vn_1}) into (\ref{Eq_des_Vn_2}) and
by using the condition that the functions $V^\pm_{n,\omega}(\phi ;
r',\phi')$ and their derivatives must be single-valued (now, we
impose periodicity in the coordinate $\phi$), we determine the
functions ${\cal A}^\pm_{n,\omega}(r',\phi')$, ${\cal
B}^\pm_{n,\omega}(r',\phi')$, ${\cal C}^\pm_{n,\omega}(r',\phi')$
and ${\cal D}^\pm_{n,\omega}(r',\phi')$ and we obtain
\begin{eqnarray}\label{Exp_des_Vn_2}
& & V^\pm_{n,\omega}(\phi ; r',\phi')= \mp
\frac{f_{\nu_{n}^\pm(\omega)}(r') / {\sqrt{r'}}}{4\pi
N^\pm_n(\omega) {\nu_{n}^\pm(\omega)} \sin [\pi \nu_n^\pm(\omega)]}
\nonumber \\
& & \times [ \Theta(\phi' -\phi) \cos [
{\nu_{n}^\pm(\omega)}(\pi-\phi'+\phi)]
\nonumber \\
& & \qquad \qquad   + \Theta(\phi -\phi') \cos [
{\nu_{n}^\pm(\omega)} (\pi +\phi'-\phi)]  ] .
\end{eqnarray}
Now, by inserting (\ref{Exp_des_Vn_2}) into (\ref{G_helm_diff}) and
by taking into account the expression (\ref{ortho_RM_2}) of the
normalization factor $N^\pm_n(\omega)$, we obtain the final result
\begin{widetext}
\begin{eqnarray}\label{G_helm_diff_expfinale_1}
& & G_\omega^{d}(r,\phi;r',\phi')=+ \frac{i
\ell}{2\pi}\sum_{n=0}^{+\infty} \frac{a_n(\omega)}{\sin [\pi
\nu_{n}^{+}(\omega)]}
\frac{f_{\nu_{n}^{+}(\omega)}(r)\;f_{\nu_{n}^{+}(\omega)}(r')}{\sqrt{rr'}}
\left\{ \Theta(\phi' -\phi) \cos [
{\nu_{n}^+(\omega)}(\pi-\phi'+\phi)] \right. \nonumber \\
&& \qquad \qquad \qquad \qquad \qquad \qquad \qquad \qquad \qquad
\qquad \qquad \qquad \qquad \qquad  \left. + \Theta(\phi -\phi')
\cos [
{\nu_{n}^+(\omega)} (\pi -\phi +\phi')]  \right\} \nonumber \\
& & \phantom{G_\omega^{d}(r,\phi;r',\phi')=} - \frac{i
\ell}{2\pi}\sum_{n=0}^{+\infty} \frac{a_n(\omega)}{\sin [\pi
\nu_{n}^{-}(\omega)]}
\frac{f_{\nu_{n}^{-}(\omega)}(r)\;f_{\nu_{n}^{-}(\omega)}(r')}{\sqrt{rr'}}
\left\{ \Theta(\phi -\phi') \cos [ {\nu_{n}^-(\omega)} (\pi
-\phi +\phi')] \right. \nonumber \\
&& \qquad \qquad \qquad \qquad \qquad \qquad \qquad \qquad \qquad
\qquad \qquad \qquad \qquad \qquad \left. + \Theta(\phi' -\phi) \cos
[ {\nu_{n}^-(\omega)}(\pi-\phi'+\phi)]  \right\} \nonumber\\
& &
\end{eqnarray}
with
\begin{equation}\label{G_helm_diff_expfinale_2}
a_n(\omega)= \frac{\Gamma\left(n+1-i\frac{\ell^2\omega}{r_h}\right)
\Gamma\left(n+2-i\frac{\ell^2\omega}{r_h}\right)}{n! (n+1)! \,
\left[\Gamma\left(1-i\frac{\ell^2\omega}{r_h}\right) \right]^2}
=\frac{\left(n+1-i\frac{\ell^2\omega}{r_h}\right)
\left[\left(1-i\frac{\ell^2\omega}{r_h}\right)_{n}\right]^2} {n!
(n+1)!}.
\end{equation}
Here the symmetry of $G_\omega^{d}(r,\phi;r',\phi')$ under the
exchange $(r,\phi) \leftrightarrow (r',\phi')$ appears explicitly.
$G_\omega^{d}(r,\phi;r',\phi')$ has been constructed for $\omega
>0$. By analytic continuation from the positive real $\omega$-axis,
it is obtained in the full complex $\omega$-plane.

Before exploiting the expression of $G_\omega^{d}(r,\phi;r',\phi')$
given by (\ref{G_helm_diff_expfinale_1}) and
(\ref{G_helm_diff_expfinale_2}) in order to recover the resonant
aspects of the BTZ black hole, it seems to us interesting to make a
digression which provide a physical interpretation of
$G_\omega^{d}(r,\phi;r',\phi')$. By using
\begin{equation}\label{Sin_explosé}
\frac{1}{\sin \pi \nu }=-2i \sum_{k=0}^{+\infty} e^{ +i\pi(2k+1)\nu}
\quad \mathrm{for} \quad \mathrm{Im} \ \nu > 0 \quad \mathrm{and}
\quad \frac{1}{\sin \pi \nu }=+2i \sum_{k=0}^{+\infty} e^{
-i\pi(2k+1)\nu} \quad \mathrm{for} \quad \mathrm{Im} \ \nu < 0,
\end{equation}
we can write
\begin{eqnarray}\label{G_helm_diff_expfinale_explosée}
& & G_\omega^{d}(r,\phi;r',\phi')=+ \frac{
\ell}{2\pi}\sum_{n=0}^{+\infty} a_n(\omega)
\frac{f_{\nu_{n}^{+}(\omega)}(r)\;f_{\nu_{n}^{+}(\omega)}(r')}{\sqrt{rr'}}
\nonumber \\
& & \qquad \times \left\{ \Theta(\phi' -\phi)
\left[\sum_{k=0}^{+\infty} e^{i[ {\nu_{n}^+(\omega)}(\phi'-\phi+k
2\pi)]} +e^{i[ {\nu_{n}^+(\omega)}(2\pi-\phi'+\phi+k 2\pi)]}\right]
+  \phi \leftrightarrow \phi'   \right\} \nonumber \\
& & \phantom{G_\omega^{d}(r,\phi;r',\phi')=} +
\frac{\ell}{2\pi}\sum_{n=0}^{+\infty} a_n(\omega)
\frac{f_{\nu_{n}^{-}(\omega)}(r)\;f_{\nu_{n}^{-}(\omega)}(r')}{\sqrt{rr'}}\nonumber \\
& & \qquad \times \left\{ \Theta(\phi -\phi')
\left[\sum_{k=0}^{+\infty} e^{-i[ {\nu_{n}^-(\omega)}(\phi-\phi'+k
2\pi)]} +e^{-i[ {\nu_{n}^-(\omega)}(2\pi-\phi+\phi'+k 2\pi)]}\right]
+ \phi' \leftrightarrow \phi    \right\}.
\end{eqnarray}
\end{widetext}
By reinstating the temporal dependance in $e^{i\omega(t-t')}$ into
Eq.~(\ref{G_helm_diff_expfinale_explosée}) [see Eq.~(\ref{GF et
Gom})], this expression provides a physical interpretation of the
diffractive Feynman propagator. It appears as a sum over $n\in
\mathbf{N}$, i.e., over all the surface waves supported by the BTZ
black hole boundary. In that sum, terms like $e^{i[ {\nu_{n}^\pm
(\omega)}(\phi'-\phi+k 2\pi)]}$ correspond to the contributions of
surface waves propagating around the black hole and the sums over
the index $k$ take into account their multiple circumnavigations.

We shall now complete this section by considering the analytic
structure of $G_\omega^{d}(r,\phi;r',\phi')$ in the complex
$\omega$-plane and, more precisely, by looking for its poles in this
plane because they correspond to the resonance frequencies of the
scalar field $\Phi$ propagating in the BTZ black hole space-time.
The only poles of $G_\omega^{d}(r,\phi;r',\phi')$ are the zeros of
the functions $\sin [\pi \nu_{n}^\pm (\omega)]$ with $n\in
\mathbf{N}$. They are therefore obtained by solving
\begin{equation} \label{CondResonance}
\nu_{n}^\pm(\omega) =m \quad \mathrm{with} \,\, m\in \mathbf{Z}
\end{equation}
which provides the exact results given by (\ref{freqcomp_QNM}) for
the QNM complex frequencies and which furthermore clarifies the
meaning of the indices $n$ and $m$ introduced to denote them: the
QNM complex frequencies are grouped into families labeled by the
indices $n\in \mathbf{N}$, each family being associated with a given
surface wave or equivalently with a Regge pole, and the members of a
given family are indexed by $m\in \mathbf{Z}$.

It is worth noting that the method we have developed in this article
should not only be regarded as a new way to calculate the QNM
complex frequencies $\omega^\pm_{mn}$. It is, above all, an approach
which permits us to physically interpret them: they appear as
Breit-Wigner-type resonances generated by the family of surface
waves supported by the BTZ black hole boundary at infinity. Indeed:

\qquad - In the immediate neighborhood of a QNM complex frequency
$\omega^\pm_{m n}$, $G_\omega^{d}(r,\phi;r',\phi')$ given by
(\ref{G_helm_diff_expfinale_1}) has the Breit-Wigner form, i.e., we
can write
\begin{equation}\label{BW}
G_\omega^{d}(r,\phi;r',\phi')  \approx \frac{{\cal N}^\pm_\omega
(r,\phi;r',\phi')}{\omega -\omega^{\pm \, (o)}_{mn} +i\Gamma^\pm
_{mn}/2}.
\end{equation}
with $\omega^{\pm \, (o)}_{mn}=\pm m/\ell$ and $\Gamma^\pm _{mn}/2
=(2r_h/\ell^2)(n+1)$. This result is a direct consequence of the
formula $\sin (\pi x) \approx (-1)^m \pi (x-m)$ for $x \to m$ with
$m \in \mathbf{Z}$.

\qquad - For a given value of $n$, a term like $1/\sin [\pi
\nu_{n}^\pm (\omega)]$ is produced by interference between the
different components of the $n$-th Regge surface wave supported by
the black hole boundary [see Eq.~(\ref{Sin_explosé}) and compare
(\ref{G_helm_diff_expfinale_1}) with
(\ref{G_helm_diff_expfinale_explosée})], each component
corresponding to a different number of circumnavigations.
Furthermore, a constructive interference between its different
components occurs when the quantity $\mathrm{Re} \, \nu_{n}^\pm
(\omega)$ coincides with an integer, i.e., for real resonance
frequencies $\omega^{\pm \, (o)}_{mn}$ obtained from the
Bohr-Sommerfeld-type quantization conditions
\begin{equation}\label{sc1}
\mathrm{Re} \,  \nu_{n}^\pm  \left(\omega^{\pm \, (o)}_{mn} \right)=
m \qquad m \in \mathbf{Z}.
\end{equation}
This equation provides again the real parts $\omega^{\pm \,
(o)}_{mn}$ of the family of QNM complex frequencies
$\omega^\pm_{mn}$ generated by the $n$-th Regge surface wave.

\section{Conclusion and perspectives}

In the present paper, by considering the Regge poles of the spinless
BTZ black hole, we have provided a new interpretation for its QNMs:
they can be considered as generated by surface waves propagating on
its boundary and the associated complex frequencies are
Breit-Wigner-type resonances. This interpretation can be easily
extended to the rotating BTZ black hole
\cite{DecaniniFolacci2009_unpublished}. In that case, the Regge
modes are associated with the Regge poles
\begin{subequations}\label{ReggeP_rotating}
\begin{eqnarray}
& & \nu_{n}^+(\omega) =+ \omega \ell +
i\frac{2(r_+-r_-)}{\ell}(n+1)  \label{ReggeP_rotating_a} \\
& & \nu_{n}^-(\omega) =- \omega \ell - i\frac{2(r_++r_-)}{\ell}(n+1)
\label{ReggeP_rotating_b}
\end{eqnarray}
\end{subequations}
with $\omega>0$ and $n\in \mathbf{N}$. Here $r_+$ and $r_-$ denote
the outer and inner horizon radii of the rotating BTZ black hole
which are linked with its mass $M$ and its angular momentum $J$ by
$M=(r^2_+ + r^2_-)/\ell^2$ and $J=2r_+r_-/\ell$. The Regge modes
correspond again to surface waves supported by the boundary of the
black hole but, now, it should be noted that they describe surface
waves with different properties. Indeed, even if they have the same
azimuthal propagation constant $|\mathrm{Re} \,
\nu_{n}^\pm(\omega)|=\omega \ell$, their attenuations are different:
the damping constant $|\mathrm{Im} \, \nu_{n}^+(\omega)|=[2(r_+
-r_-)/\ell](n+1)$ of the wave which propagates counterclockwise
around the BTZ black hole (i.e., which is in co-rotation with the
black hole) is lesser than the damping constant $|\mathrm{Im} \,
\nu_{n}^-(\omega)|=[2(r_+ + r_-)/\ell](n+1)$ of the wave which
propagates clockwise. Such a behavior leads directly to the
splitting of the QNM complex frequencies: when we insert
(\ref{ReggeP_rotating_a}) and (\ref{ReggeP_rotating_b}) into the
resonance condition (\ref{CondResonance}), we obtain the exact
results \cite{Birmingham2001}
\begin{subequations}\label{freqcomp_QNM_rotating}
\begin{eqnarray}
& & \omega_{mn}^+=+\frac{m}{\ell}-i \frac{2(r_+-r_-)}{\ell^2}(n+1)
\label{freqcomp_QNM_rotating_a} \\
& & \omega_{mn}^-=-\frac{m}{\ell}-i \frac{2(r_++r_-)}{\ell^2}(n+1)
\label{freqcomp_QNM_rotating_b}
\end{eqnarray}
\end{subequations}
with $m \in \mathbf{Z}$ and $n\in \mathbf{N}$. Now we have
$\omega_{-mn}^+ \not= \omega_{mn}^-$ and the two-fold degeneracy of
the QNM complex frequencies noted in Sec.~II is removed due to the
rotation of the BTZ black hole which induces different damping for
the surface waves propagating counterclockwise and clockwise.

Because the BTZ black hole geometry frequently appears as a factor
in the near horizon geometry of higher dimensional black holes of
string theories (see, e.g., Ref.~\cite{aharony-2000-323}), our Regge
poles analysis could be directly generalized and analogous results
could be obtained in a more general context. Similarly, it would be
interesting to explore the more general situation of quantum
corrected BTZ black holes (see, e.g., Ref.~\cite{Konoplya2004}). We
believe that all these results could be helpful in order to shed
light on AdS/CFT correspondence from a new point of view.
Unfortunately, we have been unable to make important steps in this
direction. In particular, we do not actually have at our disposal a
clear $\mathrm{CFT}_2$-interpretation of the BTZ Regge poles
analogous to the interpretation of the QNM complex frequencies
provided in
Refs.~\cite{BirminghamSachsSolodukhin2002,BirminghamSachsSolodukhin2003}.

It is finally important to recall that the boundary of the BTZ black
hole can also be considered as its photon sphere. It is therefore
quite tempting to wonder if the photon sphere of all the other black
holes might play a central role in the context of holography. In
particular, could one holographically map a quantum field theory (or
a string theory) defined on the Schwarzschild black hole of mass $M$
on a conformally invariant quantum field theory defined on its
photon sphere at $r=3M$? And more generally, could one not
systematically consider photon spheres as holographic screens?

\begin{acknowledgments}
We thank St\'ephane Ancey, Denis Bernard, Paul Gabrielli, Bruce
Jensen and Bernard Raffaelli for various discussions concerning some
of the topics considered in this article.
\end{acknowledgments}

\bibliography{RPandBTZ}

\end{document}